# Thermal-Aware Task Allocation and Scheduling for Embedded Systems

W-L. Hung, Y. Xie, N. Vijaykrishnan, M. Kandemir, and M. J. Irwin
The Pennsylvania State University, University Park, PA 16802, USA
{whung,yuanxie,vijay,kandemir,mji@cse.psu.edu}

## Abstract

*Temperature affects not only the reliability but also the performance, power, and cost of the embedded system. This paper proposes a thermal-aware task allocation and scheduling algorithm for embedded systems. The algorithm is used as a sub-routine for hardware/software co-synthesis to reduce the peak temperature and achieve a thermally even distribution while meeting real time constraints. The paper investigates both power-aware and thermal-aware approaches to task allocation and scheduling. The experimental results show that the thermal-aware approach outperforms the power-aware schemes in terms of maximal and average temperature reductions. To the best of our knowledge, this is the first task allocation and scheduling algorithm that takes temperature into consideration.*

## 1. Introduction

Traditional allocation and scheduling routines use performance or power as the design metric in Hardware/software co-synthesis [1]. As technology scales, temperature in modern high-performance VLSI circuits has moved up dramatically due to smaller feature sizes, higher packing densities and rising power consumptions. Temperature affects not only the reliability but also the performance, power, and cost of the embedded system. At sufficiently high temperatures, many failure mechanisms (such as electromigration and stress migration) are significantly accelerated, resulting in reduced system reliability; interconnect delay increases and MOS current drive capability decreases as chip temperature increases. The leakage power increases exponentially with the temperature increase; finally, the cost of cooling a hot chip increases as the hot spot temperature goes up.

Power-aware design alone is not able to address the temperature challenge, and many low-power techniques have insufficient impact on chip temperature because they do not directly target the *spatial and temporal* behavior of the operating temperature. Therefore, even though it is related to the power-aware design area, thermal-aware design itself is a distinct and important research area. In this paper, we investigate both power-aware and thermal-aware approaches for task allocation and scheduling. The experimental results show that thermal-aware approach outperforms the power-aware schemes in terms of maximal and average temperature reductions.

## 2. Tasks Allocation and Scheduling

For either platform-based or customized architecture, the task Allocation and Scheduling Procedures (ASP) is critical to get good solutions. The selection of PEs and the assignment of tasks

This work was supported in part by grants from PDG, NSF CAREER Awards 0093082 and 0093085 and GSRC.

are both guided by ASP. Our task allocation and scheduling procedure is similar to the one proposed by Xie and Wolf [1]. The ASP takes the task graph and architecture (either pre-defined platform architecture or a customized architecture generated via co-synthesis) and a target library as input, and generates the task mapping and scheduling on the target architecture. The target library stores the worst case power consumptions (WCPC) and worst case execution times (WCET) for a task executed on different PEs.

The static criticality (SC) for each task is calculated as the maximum distance from current task to the end task in a task graph. This is similar to the priority ordering in some list schedulers. The dynamic criticality (DC) calculation is based on four different factors and will be defined in section 2.1.

The traditional allocation and scheduling algorithm is effective on finding the task mapping and scheduling that satisfy the deadline requirement. However, it neglects the temperature impacts during the process. To account for this problem, we introduce power/energy aware ASP and thermal-aware ASP.

### 2.1 Power-aware allocation and scheduling

Since temperature is closely related to the power density, in power-aware allocation and scheduling, the power/energy factor is involved in the process of calculating dynamic criticality. Therefore, the *DC* equation is defined as follows:

$$DC(task_i, PE_j) = SC(task_i) - WCET(task_i, PE_j) -$$

$$max(avl.\_PE_j, ready\_task_i) - Pow$$

The first term stands for the static criticality of the $task_i$; the second term retrieves the WCET of this $task_i$ executed on $PE_j$ from the technology library, and the third term takes the maximum of $PE_j$'s available time and $task_i$'s ready time. The last term (*Pow*) captures the effect of power/energy which can be interpreted by the following three heuristics:

    Heuristic 1: minimize power consumption of current task
    Heuristic 2: minimize cumulative average power of
                processing element
    Heuristic 3: minimize energy of current task

### 2.2 Thermal-aware allocation and scheduling

The proposed thermal-aware ASP addresses the thermal issue by taking the temperature into consideration. The temperature of an embedded system depends on the power consumption of each processing element (PE), its dimension and relative location on the embedded system platform. The thermal modeling tool, HotSpot [2], is used to extract the temperature profile. Hotspot provides a simple compact model, where the heat dissipation within each PE and the heat flow among PEs are accounted for. HotSpot takes a system floorplanning and the power consumption for each function block as input, and generates accurate temperature estimation for each block.

For the thermal-aware ASP, we first pass the cumulating power consumptions of each PE along with the consuming power incurred by current scheduled task to the HotSpot. The temperatures returned from the HotSpot are averaged and then be used in calculating dynamic criticality as defined before. The newly added *Avg._Temp* substitutes out the *Pow* term and sets the goal of minimization of the average temperature. The goal also implies the reduction on the maximal temperature.

The flow of our thermal-aware co-synthesis framework is shown in Figure 1.a. The allocation and scheduling procedure executes and then activates the thermal-aware floorplanning [3] when considering assignment of a task on one specific PE. The HotSpot tool interacts with the floorplanning procedure to provide information of temperature. For the platform-based thermal-aware design, the target architecture and the task graph are given, and the HotSpot is activated by the modified ASP with thermal inquires. This flow is depicted in Figure 1.b.

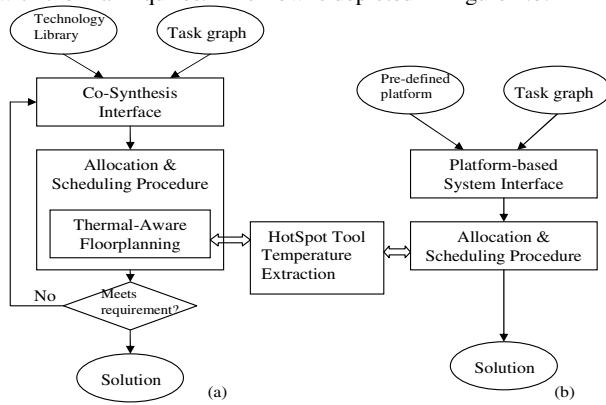

**Figure 1. The flows of the thermal-aware co-synthesis framework and thermal-aware platform-based system design**

## 3. Experimental Results

The first experiment we conduct is to compare the temperature differences from different power heuristics when using the co-synthesis to decide the selection of PEs and when using the platform-based architecture (using four identical PEs). The experimental results are shown in Table 1. The three columns under the *co-synthesis* are the results of the traditional co-synthesis work, while the other three columns represent the results from the platform-based target architecture.

The very first row out of four rows' groups indicates the characteristics of each benchmark and is the baseline case that does not take the power into consideration. The following three rows represent three power heuristics. As can be seen from the table, when considering power only, the third power heuristic outperforms the other two heuristics and the baseline approach. This result indicates that minimizing the energy of a task executed on one specific PE achieves the best temperature result among all three heuristics. Thus, the third power heuristic will be used in the following experiments.

The second experiment is to demonstrate the effectiveness of our thermal-aware approach in terms of lowering the peak and the average temperatures. We take the best results of customized architecture and platform-based architecture from the first experiment for comparison. The power-aware and thermal-aware customized architecture comparison is shown in Table 2. From the results, the customized architecture with thermal-aware approach demonstrates that it can effectively reduce the total average temperature reduction by 10.9 $^o$C and 6.95 $^o$C for the maximal and the average, respectively. This result indicates that observing the average temperature of all using PEs while doing task scheduling is beneficial to control the temperature of an embedded system.

**Table 1. The comparisons of different power heuristics under co-synthesis arch. and platform-based target arch.**

| name/task/edge/ deadline | co-synthesis | | | Platform-based Arch. | | |
|---|---|---|---|---|---|---|
| | Total Pow. | Max Temp. | Avg Temp. | Total Pow. | Max Temp. | Avg Temp. |
| **Bm1/19/19/790** | 16.60 | 118.18 | 106.32 | 11.91 | 100.59 | 81.03 |
| Heuristic 1 | 16.14 | 121.7 | 109.29 | 10.40 | 85.88 | 75.58 |
| Heuristic 2 | 16.60 | 118.18 | 106.32 | 12.60 | 107.16 | 82.78 |
| Heuristic 3 | 15.56 | 113.29 | 104.49 | 10.40 | 85.88 | 75.58 |
| **Bm2/35/40/1500** | 29.47 | 121.44 | 110.22 | 24.48 | 114.33 | 101.04 |
| Heuristic 1 | 28.55 | 115.21 | 107.55 | 23.36 | 107.63 | 98.21 |
| Heuristic 2 | 29.47 | 121.44 | 110.22 | 24.90 | 113.31 | 99.96 |
| Heuristic 3 | 28.27 | 112.82 | 105.42 | 24.09 | 106.63 | 97.4 |
| **Bm3/39/43/1650** | 28.84 | 113.58 | 101.76 | 26.88 | 113.81 | 98.47 |
| Heuristic 1 | 27.75 | 110.33 | 100.46 | 26.1 | 106.63 | 96.74 |
| Heuristic 2 | 29.35 | 110.49 | 100.6 | 26.88 | 113.81 | 98.47 |
| Heuristic 3 | 28.20 | 109.96 | 100.15 | 25.20 | 103.95 | 94.69 |
| **Bm4/51/60/2000** | 44.99 | 122.09 | 111.14 | 42.35 | 106.54 | 97.05 |
| Heuristic 1 | 46.99 | 122.28 | 111.53 | 40.33 | 100.61 | 89.74 |
| Heuristic 2 | 44.99 | 117.86 | 111.13 | 42.35 | 106.54 | 91.62 |
| Heuristic 3 | 43.34 | 118.68 | 109.87 | 41.64 | 100.42 | 89.24 |

As for the platform-based architecture, the proposed thermal-aware approach outperforms the power-aware approach in both temperature attempts. As shown in Table 3, under thermal-aware approach, both of the maximal and average temperatures are lower than those of in the corresponding power-aware approach and approximately by 9.75 $^o$C and 5.02 $^o$C, respectively.

**Table 2. The temperature comparisons of the power-aware and the thermal-aware approaches on co-synthesis architecture.**

| Bechmark | Power-aware co-synthesis | | | Thermal-aware co-synthesis | | |
|---|---|---|---|---|---|---|
| | Total Pow. | Max Temp. | Avg Temp | Total Pow. | Max Temp. | Avg Temp. |
| Bm1 | 15.56 | 113.29 | 104.49 | 12.48 | 87.11 | 86.13 |
| Bm2 | 28.27 | 112.82 | 105.42 | 24.64 | 106.38 | 99.84 |
| Bm3 | 28.2 | 109.96 | 100.15 | 26.51 | 102.08 | 96.28 |
| Bm4 | 43.34 | 118.68 | 109.87 | 42.41 | 106.32 | 102.48 |

The results from Table 2 and 3 indicate that with the platform-based architecture, the thermal ASP can balance the workloads of all PEs, and thus delivery a lower peak and average temperatures task mapping than that of in customized architecture.

**Table 3. The temperature comparisons of the power-aware and the thermal-aware approaches on platform-based architecture.**

| Bechmark | Power-aware platform Arch. | | | Thermal-aware platform Arch. | | |
|---|---|---|---|---|---|---|
| | Total Pow. | Max Temp. | Avg Temp | Total Pow. | Max Temp. | Avg Temp. |
| Bm1 | 10.40 | 85.88 | 75.58 | 6.37 | 65.71 | 61.16 |
| Bm2 | 24.09 | 106.63 | 97.40 | 22.37 | 96.33 | 93.47 |
| Bm3 | 25.20 | 103.95 | 94.69 | 24.98 | 103.03 | 94.59 |
| Bm4 | 41.64 | 100.42 | 89.24 | 38.54 | 94.85 | 85.76 |

## References

[1] Yuan Xie and Wayne Wolf, "*Allocation and scheduling of conditional task graph in hardware/software co-synthesis*", DATE 2001.
[2] K. Skadron, T. Abdelzaher, and M. Stan, "*Control-Theoretic Techniques and Thermal-RC Modeling for Accurate and Localized Dynamic Thermal Management*", HPCA 2002.
[3] W-L. Hung Y. Xie, N. Vijaykrishnan, C. Addo-Quaye, T. Theocharides, and M. J. Irwin, "*Thermal-Aware Floorplanning Using Genetic Algorithms*", ISQED 2005.